\documentstyle[PASJadd,psfig]{PASJ95}
\draft
\begin{document}
\setcounter{page}{1}
\title{ASCA Temperature Maps of Three Clusters of Galaxies Abell~1060, AWM~7, 
and the Centaurus Cluster}

\author{T.~{\sc Furusho},$^1$ N. Y.~{\sc Yamasaki},$^1$ T.~{\sc
    Ohashi},$^1$ R.~{\sc Shibata},$^2$ T.~{\sc Kagei},$^1$ Y.~{\sc
    Ishisaki},$^1$ \\ K.~{\sc Kikuchi},$^3$ H.~{\sc Ezawa},$^4$ and
  Y.~{\sc Ikebe}$^5$ \\ [12pt] $^1${\it Department of Physics, Tokyo
    Metropolitan University, 1-1 Minami-Ohsawa, Hachioji, Tokyo
    192-0397}\\ {\it E-mail(TF): furusho@phys.metro-u.ac.jp}\\ $^2${\it
    Institute of Space and Astronautical Science, 3-1-1 Yoshinodai,
    Sagamihara, Kanagawa 229-8510}\\ $^3${\it Tsukuba Space Center,
    National Space Development Agency (NASDA), 2-1-1 Sengen, Ibaraki
    305-8505}\\ $^4${\it Nobeyama Radio Observatory, National
    Astronomical Observatory, 462-2 Minamimaki, Minamisaku, Nagano
    384-1305}\\ $^5${\it Max-Planck-Institute f\"ur extraterrestrische
    Physik, Postfach 1603, D-85740, Garching, Germany} }

\abst{ We present two-dimensional temperature maps of three bright
  clusters of galaxies Abell 1060, AWM~7, and the Centaurus cluster,
  based on multi-pointing observations with the ASCA GIS\@.  The
  temperatures are derived from hardness ratios by taking into account
  the XRT response. For the Centaurus cluster, we subtracted the
  central cool component using the previous ASCA and ROSAT results,
  and the metallicity gradients observed in AWM~7 and the Centaurus
  cluster were included in deriving the temperatures.  The
  intracluster medium in Abell~1060 and AWM~7 is almost isothermal from the
  center to outer regions with a temperature of $3.3$ and $3.9$ keV,
  respectively.  The Centaurus cluster exhibits remarkable hot regions
  within about $30'$ from the cluster center showing a temperature
  increase of $+0.8$ keV from the surrounding level of 3.5 keV, and
  outer cool regions with lower temperatures by $-1.3$ keV\@.  These
  results imply that a strong merger has occurred in the Centaurus in
  the recent $2-3$ Gyr, and the central cool component has survived
  it. In contrast, the gas in Abell~1060 was well-mixed in an early
  period, which probably has prevented the development of the central
  cool component. In AWM~7, mixing of the gas should have occurred in
  a period earlier than the epoch of metal enrichment.}

\kword{ galaxies: clusters: individual (Abell~1060, AWM~7, Centaurus Cluster)
        --- galaxies: intergalactic medium
        --- X-rays: galaxies}

\maketitle \thispagestyle{headings}

\section{Introduction} 

Clusters of galaxies are filled with a diffuse, X-ray emitting hot
plasma with temperatures typically a few $\times 10^7$ K\@.  The
primary energy source of the intracluster medium (ICM) is considered
to be the kinetic energy released through the gas infall into the
gravitational potential. The heating mechanism is, however, not yet
fully understood. Numerical simulations predict that shock heatings
are caused by mergers of smaller groups and subclusters, and recent
observations provide evidences as temperature structures.  Mergers are
considered to be the common phenomenon in the cluster evolution, which
can be studied by the cluster morphologies and the temperature
structures.

Another common feature in clusters of galaxies is the existence of a
cool gas component in the cluster center.  The presence and intensity
of the cool components are recognized either from central narrow peaks
in the surface brightness profiles or from X-ray spectra. The origin
of the cool component has often been interpreted in terms of cooling
flow models (e.g. Fabian 1994), in which dense gas in the cluster
cores has radiative cooling times comparable to or shorter than the
Hubble time.  The cooling flow model requires a multi-phase
temperature structure, whereas extensive cluster studies from ASCA
spectroscopy have shown the co-existence of hot ($kT\gtsim 3$ keV) and
cool ($kT \sim 1$ keV) components in the central peaks. Strong
connection between the cool component and the cD galaxy has been
pointed out (Ikebe et al.\ 1999), and it is important to know when and
how these cool components have been created in the course of cluster
evolution.

Clusters with irregular morphologies and hot clusters with no central
cool components often show significant temperature structures which
are considered to be evidences of ongoing mergers (e.g. Markevitch et
al.\ 1998).  In particular, two-dimensional temperature maps of nearby
clusters show detailed features with ASCA and various remarkable
structures have been found.  For example, hot regions larger than 100
kpc were discovered from the Coma (Honda et al.\ 1996; Donnelly et
al.\ 1999; Watanabe et al.\ 1999) and the Virgo (Kikuchi et al.\ 2000;
Shibata et al.\ 2001) clusters.  In the Virgo cluster, the hot region
with $kT \sim 4$ keV was twice higher than the surrounding regions.
Under a merger scenario, the subcluster falls into the gravitational
potential whose depth is characterized by the ICM temperature.  If the
infall occurs with the free-fall velocity which is about the same as
the random velocity of the gas particles in the ICM, then the
instantaneous rise of temperature after the collision is roughly by a
factor of 2.  This would be the maximum expected temperature, since
the heat will quickly dissipate into the surrounding ICM\@.

To obtain a consistent view of the evolution of clusters with
different sizes and morphologies, we need to look into temperature
structures in symmetric clusters which show cool components.
Numerical simulations show that temperature structures persist over
several Gyr even after the surface brightness profile has been almost
completely smoothed out (e.g.\ Schindler and M\"uller 1993).  The
Centaurus cluster, which has a strong central cool component (Fukazawa
et al.\ 1994; Ikebe et al.\ 1999), shows a hot region at a few 100 kpc
from the center (Churazov et al.\ 1999). Therefore, sensitive studies
of the temperature distribution for a larger cluster sample and to
their outer regions would bring us rich information.

This paper presents temperature maps of Abell 1060 (A1060 for short),
AWM~7, and the Centaurus cluster (Abell 3526), based on the
multi-pointing observations with ASCA (Tanaka et al.\ 1994). These
clusters are all poor, nearby ($z=0.0104-0.0172$), and bright clusters
with the ICM temperatures $kT = 3-4$ keV\@. The closeness of these
systems allows ASCA to resolve the temperature structures less than 100
kpc. The clusters are markedly different regarding the strength of the
central cool component. The Centaurus cluster shows a very sharp central
peak and strong cool component (Fukazawa et al.\ 1994; Allen and Fabian
1994; Ikebe et al.\ 1999).  Strong metal concentration up to about 1.5
solar is seen, and a giant elliptical galaxy, NGC4696, exists as the cD galaxy. In contrast, A1060 does not show such a
narrow peak and the X-ray surface brightness agrees well with a single
$\beta$ model from the center to a radius of $1h_{50}^{\ -1}$ Mpc.  The
metal abundance is flat with radius and the central cool component is
very weak or absent (Tamura et al.\ 1996, 2000). These properties are
typical of non-cD clusters. Another cluster AWM~7 has a cD galaxy, NGC
1129, but the central X-ray peak is not as narrow nor intense as
Centaurus. The metal concentration and the central cool emission are
definitely weaker or more extended than the Centaurus (Neumann and
B\"{o}hringer\ 1995; Xu et al.\ 1997; Ezawa et al.\ 1997). Therefore,
AWM~7 falls in the middle of Centaurus and A1060 regarding the concentration
of the cool component.  For example, ratios of the cool-component luminosity
to the total one are $\sim 10\%$, 6\%, and $<3\%$ for Centaurus, AWM~7,
and A1060, respectively (Tamura et al.\ 1997).  This difference may
reflect the past history of the cluster evolution, such as a recent
occurrence of mergers which might have disrupted a central cool component.
Comparison of temperature maps among the three clusters is interesting
in examining the connection between temperature structures and cD
dominance.

Throughout this paper, we assume $H_0 = 50$ km s$^{-1}$ Mpc$^{-1}$ and
$q_0 = 0.5$. The solar number abundance of Fe relative to H is taken
as $4.68 \times 10^{-5}$ (Anders and Grevesse\ 1989).

\section{Observations} 

\subsection{ASCA observations and Data reduction}

Table~1 shows the log of ASCA observations.  The numbers of the pointings
 are 7, 6, and 11 for A1060, AWM~7, and Centaurus,
respectively. Two pointings for A1060 have a very small separation
angle ($\sim 2'$).  We selected the GIS data with a cut-off rigidity
$> 8$ GV and the telescope elevation angle $> 10^\circ$ and $> 5^\circ$
from the earth rim for day and night, respectively (see Ohashi et al.\
1996 and Makishima et al.\ 1996 for the details of the GIS system).
This paper deals with only the GIS results, since the field of view of
the SIS ($11'\times11'$) was unable to cover the whole cluster fields
and not suitable for the present purpose.

Some of the spectra of A1060, taken in December 1999, showed
significant excess emission in the energy range below 1 keV\@.  The
excess is thought to be due to scattered solar X-rays from the earth
atmosphere, because a medium class solar flare occurred during the ASCA
observation. By setting the elevation angle from the day earth to be
$> 20^\circ$, this soft excess was eliminated. This condition was
applied to all the A1060 data. The flare-like events due to the
fluctuation of the particle background (Ishisaki\ 1996) were screened
out, and the time periods after the attitude maneuver with unsettled
pointing directions were excluded.  The non X-ray and the cosmic X-ray
backgrounds were estimated from the archival data taken during
1993--1994 (Ikebe\ 1995).  Since all targets were observed during
1993--1999, we took into account the slow change in the intensity of
the GIS non X-ray background (Ishisaki et al.\ 1997, ASCA News No.\ 5,
26).  The long-term variation of the background count rate until 2000
indicated a factor of $\sim 1.2$ maximum around 1998 and an rms
variation $\sim 7\%$.

The GIS2 images of the 3 clusters in the 0.7 -- 7 keV band are shown
in figures~1a--c. The circles show the GIS field of views with a
radius of $22'$. The images were smoothed by a Gaussian function with
$\sigma = 1'$ after subtraction of the non X-ray and the cosmic X-ray
backgrounds and exposure correction.  In the A1060 image in figure~1a,
the source at $29'$ north-east of the cluster center is a group of
galaxies, HCG~48 ($z=0.0094$). Other clusters show no obvious
substructures or intense point sources as bright as HCG~48.

\subsection{ROSAT observations}
 
For spatially extended sources, an analysis of the ASCA data needs
proper consideration of the complicated response of the X-ray
telescope (XRT; Serlemitsos et al. 1995) as described in the next
section. To take these effects into account, we simulated the GIS
observations using the two-dimensional ROSAT PSPC data in the 0.5 -- 2.0 keV band as
input images. The higher angular resolution than ASCA by several
factors would allow such a substitution.

The ROSAT PSPC observed A1060 and AWM~7 on January 1 in 1992 and
January 28 in 1992, respectively. The single pointings of the PSPC
completely cover the GIS observed regions for these 2 clusters. For
the Centaurus cluster, 5 PSPC pointings were conducted on February 2
in 1992 (the central pointing), July 25 in 1992 and January 16 -- 25 in
1993 (the outer 4 pointings).

The data selection was carried out by looking into the time variation
of the PSPC count rate in different pulse-height channels.
Subtraction of non-cosmic background and correction for the vignetting
were performed using the Extended Source Analysis Software package
(Snowden et al.\ 1994).  Assuming the axial symmetry, we fit the
surface brightness profile in the energy range of 0.5 -- 2.0 keV with a
single $\beta$ model (double $\beta$ model for the Centaurus) added
onto a constant level which corresponds to the cosmic X-ray
background.  The obtained $\beta$-model parameters are consistent with
the previous results for A1060 by Tamura et al.\ (2000), AWM~7 by
Neumann and B\"{o}hringer\ (1995), and Centaurus by Ikebe et al.\
(1999), respectively. 
Following this consistency check, we smoothed the two-dimensional PSPC 
images by a Gaussian function in order to suppress the statistical fluctuation. The gaussian width depended on the intensity: $\sigma=60''$
and $15''$ for the intensity levels $< 2$ and 2 -- 10 cts pix$^{-1}$,
respectively, and no smoothing for $> 10$ cts pix$^{-1}$. This
procedure preserves the central sharpness of the images.

\section{Analysis}

\subsection{Hardness ratio analysis}

We estimated temperatures from hardness ratios ({\it HR}) rather than
spectral fits, in order to maximize the spatial resolution of
temperature maps.  The energy division of the 2 energy bands was
chosen so as to minimize the error of the resultant temperature for a
temperature range of 3 -- 4 keV\@.  By evaluating ${\it HR} - kT$ relation for
different energy divisions, we chose a common dividing energy of 2 keV
for all 3 clusters. The definition of the hardness ratio is ${\it HR} =
H/S$, where $H$ and $S$ are the count rates in 2 -- 7 keV and 0.7 -- 2
keV, respectively.  The actual {\it HR} values are between 0.6 and 0.9 for
all clusters.

The {\it HR}s and their statistical errors were calculated from the
background subtracted data which were combined into a large image for
each cluster from the ASCA multi-pointing observations. The {\it HR}s
were converted to the temperature based on the ${\it HR} - kT$
relation.  We assumed the incident spectra to be absorbed MEKAL models
based on the previous ASCA results. The spectral parameters are shown
in table~2. The same model was also used for the ray-tracing
simulations.  The parameters for A1060 were taken from Tamura et
al. (2000). The $N_{\rm H}$ value is a little smaller than the
Galactic value of $5\times10^{20}$ cm$^{-2}$, and the metal abundance
was fixed at $Z=0.31$ solar.  In AWM~7, the metallicity gradient was
found by Ezawa et al.\ (1997).  We included the effect in the ${\it
HR} - kT$ relation and the ray-tracing simulation, assuming the
abundance curve given in table 1 of Ezawa et al.\ (1997). The best-fit
$N_{\rm H}$ value was consistent with the Galactic value.  For the
Centaurus cluster, the metallicity gradient observed by Ikebe et al.\
(1999; equation~(5)) was used in our analysis. The effect of the
abundance variation on temperature is less than $\sim 0.13$ keV in the
Centaurus case.  The best-fit $N_{\rm H}$ values obtained by Ikebe et
al. (1999) differed with spectral models, but were consistent with the
Galactic value (Stark et al. 1992). We therefore adopted the Galactic
value here.

\subsection{Correction on XRT Response}

The XRT on-board ASCA has three main difficulties in performing
spatially resolved spectral studies for sources which are extended
more than the field of view: 1) energy-dependent decline of the
effective area as a function of an off-axis angle from the optical
axis (energy-dependent vignetting), 2) stray light from outside of the
field of view (mainly from the bright cluster core in the present
case), and 3) energy-dependence of the point spread
function (PSF).

The first effect is dominant for observations of extended sources. The
effective area of XRT at an off-axis angle of $20'$ from the optical
axis decreases to $\sim 25\%$ and 10\% for 2 and 8 keV, respectively.
This difference, on the one hand, causes an underestimate of the {\it
HR} value in the field edge, and complicates the data treatment, on
the other hand, for overlapping regions of the multi-pointings even if
the cluster is completely isothermal. Therefore, we ran ray-tracing
simulations for isothermal clusters based on the PSPC image to correct
for these effects.

Second, the stray light mixes the X-ray fluxes from different regions.
This effect can be examined also with the ray-tracing simulation by
tracing back the detected photons to their sky origins.  Figure~2
shows fractions of the photon origins in 7 annuli with $r=0'-4'-8'-12'-20'-28'-36'-44'$. The fractions are calculated from the simulated image combined with all pointings for 3 clusters. 
The top curve shows the fraction coming
from the pointed sky region, and the middle curve indicates
contaminating fraction from brighter inner regions for each
radius. The bottom curve is the contamination from outer regions. 
For the innermost region within a radius of $4'$, 80 -- 90\% of the
detected photons come from the corresponding sky region. In the outer
regions $r > 10'$, the fraction of this ``direct'' component reduces
to 50 -- 60\% and remains to be about the same to a radius of $40'$. The
contaminating flux mostly comes from inner regions, making the radial
temperature distribution smoother than the true structure.

Regarding the last effect, the PSF always produces image distortions and
has tails which depend on off-axis angles.  In particular, the tail of
the PSF is more extended in high energies than in lower energies. This
energy-dependent effect is included in our simulation which reproduces
the tail spectrum with 10\% accuracy (Kunieda et al.\ 1995, ASCA News 
No.\ 3, 3). This effect becomes significant
at an offset angle $r\sim6'$, where the source flux drops to about 1/100.
Therefore, this PSF effect is not a significant problem in the present
analysis.

Examples of the {\it HR}s are shown in figure~3, in which observed
{\it HR} profiles in an azimuthal direction are compared with
simulated values for a radius of $10'-20'$ for three clusters. The
error bars of the data are 90\% statistical errors.  For the simulated
clusters, we assume an isothermal ICM with the average temperature for
the hot component based on the previous ASCA studies as described
before. The assumed spectral parameters are listed in table~2. Because
of the XRT response, even the simulated {\it HR}s vary by about 5\%
from position to position.  The {\it HR}s are in good agreement with
the isothermal cluster in the case of A1060, while the Centaurus
clearly shows significant deviations.  We note that the isothermal
models for the simulations also included the abundance gradients for
AWM~7 and Centaurus as described in the previous subsection.

By taking the ratio of {\it HR}s between the data and the isothermal
simulation for each cell of the image, the temperature maps were
produced. The ratio in each cell gives a correction factor to the {\it
HR}, and the corrected {\it HR} values and errors were then converted
to the temperatures using an appropriate ${\it HR}-kT$ curve for the
assumed metal abundance and $N_{\rm H}$\@.  Further details concerning
the temperature estimation and systematic errors are described later
in section 5.

\subsection{Temperature maps}

Based on the hardness ratio analysis, we produced two-dimensional
temperature maps (see figure~4).  The images
were binned into a cell of $5' \times 5'$ for A1060 and AWM~7, and
$2.5' \times 2.5'$ for Centaurus, since the Centaurus is much brighter
than the others. These angular scales correspond to actual sizes of
$\sim 100$, 150, and 50 kpc for A1060, AWM~7, and Centaurus,
respectively. Furthermore, to suppress the fluctuation, we took
an intensity-weighted average for $4 \times 4$ cells ($20' \times
20'$ for A1060 and AWM~7, $10'\times 10'$ for Centaurus). Therefore,
the color-coded maps show running means of the temperature and sharp
structures are suppressed. These running means were only adopted in
the color representation in figure~4, and quantitative analysis was
carried out on individual data.  The temperature maps are shown for a
spatial range where statistical errors in the $4 \times 4$ cells are
less than 10\% for A1060 and AWM~7, and less than 15\% for Centaurus,
respectively.

Since the color-coded maps give no information about the error, we
present the temperature values for 7 annular regions in figure~5. The
dividing radii are $r=0'-2'-4'-8'-12'-20'-28'-36'$, and the data
points are mutually independent (no running means). Each ring at
$r>2'$ is equally divided to 8 azimuthal sections with an opening
angle of $45^\circ$ each, but 4 azimuthal sections with $90^\circ$ for
the $2'-4'$ ring.  The errors indicate 90\% statistical errors.

To obtain detailed spectral information for
selected regions, we also carried out spectral fits.  Observed
properties of individual clusters are described in detail below.

\section{Results} 

\subsection{A1060} 

The color-coded map in figure~4a 
shows that the temperature variation in A1060 is less than 15\% of the
average value in the whole cluster. The azimuthally averaged
temperatures are shown in figure~5a. Fitting a model of constant value
to the data in figure~5a gives the mean temperature of $3.26 \pm 0.06$
keV with $\chi^2/\nu$= 57.2/44.  The error is the variance of mean at
$\Delta\chi^2=2.706$ by the statistical errors only.  The mean
temperature is consistent with the previous results of $3.3 \pm 0.1$ keV
from spectral fitting within the central $5'$ by Tamura et al.\ (1996,
2000).

Based on the temperature map, we accumulated the pulse-height spectra
for selected regions. The south-east region (region \#2 indicated in
figure~6) suggests slightly higher temperature compared with the
region \#1, whose significance should be confirmed.  We assumed an
absorbed MEKAL model and the response for an isothermal cluster, which
do not require a priori knowledge of the temperature distribution (as
in Honda et al.\ 1996). The spectra and the best-fit models are shown
in figure~6, and the ratios between data and the best-fit model for
the medium-temperature region \#1 (``med'') are shown in the bottom
panel. The normalizations were arbitrarily scaled to help visual
comparison. Results of the spectral fits are summarized in
table~3. The 90\% error ranges of the temperatures and metal
abundances between \#1 and \#2 overlap with each other.  Since no
other regions suggest a larger deviation in temperature, we conclude
that the ICM in A1060 is consistent to be isothermal.

\subsection{AWM~7}

Figure~4b shows the temperature map of AWM~7\@.  The temperature
variation in the whole cluster is again less than 15\%.  The central
region within a radius of $20'$ is very isothermal, but there are some
structures near the north and south edges.  In figure~5b, azimuthally
averaged temperatures are shown with the mean temperature of $3.82 \pm
0.06$ keV from a constant-temperature fit with $\chi^2/\nu= 55.5/44$.
The PSPC result indicates a temperature drop due to the cool component 
in the central $r<2'$ region (Neumann and B\"ohringer 1995), however this
cool component is very weak in the GIS data and the temperature after 
averaging over a $20'\times20'$ region shows almost no signature. 
If this central region is excluded, the reduced
chi-square becomes significantly smaller, $\chi^2/\nu= 42.1/43$ with
the mean temperature of $3.85 \pm 0.06$ keV. Both of the mean
temperatures agree with the cluster average of $3.9 \pm 0.2$ keV with
the central pointing obtained by Markevitch and Vikhlinin (1997). They
also derived a two-dimensional temperature map for the central pointing
and found it isothermal. 

The temperature map in figure~4b shows cool and hot features at
$\sim20'$ north and south of the cluster center.  We also fit the
spectra for these regions (regions \#1 and \#2 in figure~7) with
absorbed MEKAL models, and the data, best-fit models, and ratios to
the ``cool'' model are shown in figure~7.  As shown in table~3, the
best-fit temperatures are $3.6^{+0.6}_{-0.4}$ keV for the north and
$4.6^{+1.2}_{-0.7}$ keV for the south regions.  The temperature
difference is larger than the 90\% error but less than the 95\% limit,
so this feature has a marginal statistical significance.

\subsection{Centaurus Cluster} 

\subsubsection*{\bf Temperature map}

Figure~4c shows the temperature map of the Centaurus cluster without
any correction to the central cool component. The central region
within $r < 30'$ shows a remarkable hot region $\sim 15'$ south east
from the center, which was already reported by Churazov et al.\
(1999). 
However, since no corrections have been made for the central cool
component, the inner envelope of the reported hot region may well be
the outer boundary of the cool component. This point needs to be
examined. Another interesting new feature is the extended cool regions
at $40'$ north west and north east, indicating a temperature close to
2 keV\@.  Figure~5c shows the azimuthally averaged temperatures. By
fitting a constant model to the data of figure~5c, isothermal model
was rejected at the 99.99\% confidence with the mean temperature of
$3.15 \pm 0.03$ keV (reduced chi-square of $\chi^2/\nu= 1110/44$).

\subsubsection*{\bf Correction for the Cool Component}

Since our main interest is the structure of the hot component, we have
subtracted the cool component from the data assuming the model derived
from Ikebe et al.\ (1999), who analyzed PSPC and GIS data jointly. The
model gives radial profiles of the emission measure and filling factor
with a cut off radius of $R_{\rm cut}=6'$ assuming a constant
temperature, $kT_{\rm cool} = 1.4$ keV, for the cool component. Using
this model and the ray-tracing simulation, we can estimate the radial
count-rate profiles for the hard and soft bands, which were simply
subtracted from the data.  The central intensity of the cool component
was estimated here from the ratio of the emission integrals of the two
components given in table~3 of Ikebe et al.\ (1999), which showed the
best-fit parameters for the spectrum within the central $3'$ region
fitted with a 2-temperature MEKAL model.  Note that the number of
photons subtracted as the cool component within the central $6'$ was
about a quarter of the total.

Figure~4d shows the resultant temperature map for the hot component
only. The difference between figure~4c and 4d mainly appears within
the central $\sim 8'$ region as expected from the model profile of the
cool component.  The cluster center is fairly hot, and the obtained
temperature ($4.3 \pm 0.4$ keV) is close to the level in the
south-east hot region.  This temperature is consistent with the result
($4.2^{+0.8}_{-0.3}$ keV) by Ikebe et al.\ (1999). The azimuthally
averaged temperatures are shown in figure~5d.  The mean temperature
for the data in figure~5d increases by 0.7 keV to $3.88 \pm 0.05$
keV\@.  Again, the isothermal model was rejected with 99.99\%
confidence, though the reduced chi-square value became considerably
smaller to $\chi^2/\nu= 125.8/44$.

Systematic errors in the subtraction process of the cool component
need to be examined here. To avoid complications due to model
parameters, we simply varied the cool-component intensity by $\pm
10\%$, by fixing its temperature and radial profile.  This is an
overestimation since the PSPC data can constrain the cool-component
intensity with an accuracy of a few \%. The 10\% change causes a
change of the hot-component temperature by 7\% in the sense that
stronger cool emission gives hotter temperature. Since the systematic
error is comparable to the statistical error, we can say that the
features in the temperature map has an accuracy of about 10\%.

\subsubsection*{\bf Spectra}

To evaluate the significance of the hot and cool temperatures,
pulse-height spectra for 3 selected regions were compared. The
selected regions, their spectra, and the ratios of each data to the
best-fit model for \#1 are shown in figure~8, and results of spectral
fits are shown in table~3. The region \#2 in the south-east includes
the hot region reported by Churazov et al.\ (1999), and the obtained
temperature of $4.3 \pm 0.2$ keV is consistent with their result of
$4.4 \pm 0.2$ keV\@. The region \#1 is at nearly symmetric to the hot
region with respect to the cluster center, which should reduce the
systematic effects in the temperature comparison. The obtained
temperature of $3.5 \pm 0.1$ keV is the typical value excluding the
hot and cool regions.  At $30'$ north-west, a cool region is seen in
the temperature map (region \#3).  The spectral fit shows the
temperature to be $2.2 ^{+0.4}_{-0.3}$ keV, which is significantly
lower than the mean temperature.

\section{Systematic errors} 

To confirm the significance of the temperature results, we need to look
into systematic errors in the temperature determination process. There
are 3 major origins to be considered: fluctuation of the cosmic (CXB)
and non X-ray (NXB) backgrounds, error in the response function, and
error in the assumed surface brightness profile.

\subsection{Background uncertainty}

The influence of the background fluctuation depends on the signal
intensity, and outer regions of the cluster are considered to be more
affected. Since both CXB and NXB have harder spectra than the observed
clusters, their fluctuation can have a significant effect on the
temperature results. We have studied this with a simulation by changing
the background intensity. The fluctuation amplitude was chosen to be
10\% for CXB and 7\% for NXB, respectively, which are the typical
values for the field to field variation, based on the previous studies
(Ishisaki et al.\ 1997, ASCA News No.\ 5, 26). 

Assuming the highest ($+10\%$ CXB and $+7\%$ NXB) and the lowest
($-10\%$ and $-7\%$) background levels, we found that the temperatures
at $r> 30'$ of A1060 changes by $-0.2$ and $+0.3$ keV on the average.
The largest change in a single region ($5'\times 5'$) was $-0.6$ and
$+0.8$ keV\@. Since other clusters have higher surface brightness than
A1060 in the outer regions, we may take the systematic error of the
temperature to be about 0.3 keV in the outermost
regions and significantly smaller in the inner regions.

\subsection{XRT response}

Here we deal with the systematic errors due to the energy-dependence of
the PSF and the effective area, and the stray light. 
The energy-dependent tail of the PSF comes the largest effect on the {\it HR}
at $\sim6'$ from the cluster center, where the tail is produced by the bright
center. Figure 2 shows that the flux contamination at $r=4'-8'$ from the inner
region is 25--35\% of the detected flux. Combining this with the 10\% 
systematic error in our simulation, the effect on the {\it HR} should be
smaller than 3.5\%, which corresponds to a temperature error of about 5\%.

As for the error on the effective area, offset
observations of the Crab nebula with the GIS showed the maximum
fluctuation of photon index to be 2\% for off-axis angles smaller than
$22'$ (Fukazawa et al.\ 1997, ASCA News No.\ 5, 3). This corresponds to a
{\it HR} error of 5.3\% and results the temperature error of $\ltsim
8$\%\@. We have also performed several isothermal simulations for A1060
with temperatures between 2.7 keV and 3.6 keV. The ${\it HR}-kT$ relation 
at a large offset angle keeps an accuracy of 0.1 keV within a radius of $40'$.

The stray light of the XRT was studied by Ishisaki (1996) based on the
Crab observations. The intensity of the stray light predicted by the
ray-tracing simulation was shown to have an error of
$15-20\%$. Figure~2 shows that fraction of the stray light, coming
from offset angles $> 40'$, is at most 50\% for all observed regions,
causing an intensity error of about 10\%. If all of this intensity
falls in the hard band, we would observe 10\% higher HR value than the
true case. The spectrum of the stray light becomes harder for offset
angles $\sim 30'$ and then softer for larger angles, but the spectral
change is not very drastic.  Considering this spectral behavior, we
may assume that the systematic error of the {\it HR} is $\pm 5\%$ in the
worst case.  The change of {\it HR} value by 5\%, causes the temperature
change by about 10\% for $kT = 3-4$ keV\@. Therefore, 10\% of
temperature is the conservative systematic error arising from our
response estimation.

\subsection{Surface brightness profile}

Since the incident emission-measure profiles in calculating simulated
clusters and response functions were based on the PSPC data, the
difference in the sensitive energy ranges between PSPC and GIS might
have caused some error in the present temperature results. The
conversion factor from the PSPC count rate to emission measure varies
by about 10\% for a temperature range of 3--5 keV (B\"ohringer 1994).
To investigate the systematic error due to the surface brightness
profile, we first approximated the PSPC image with a $\beta$ model and
modified the model parameters. The two $\beta$ models compared here
are $\beta=0.54$ and 0.61, with a common core radius of
 $3.9'$. These parameters are given in table~2 of Tamura et al.\
(2000); the core radius and the lower $\beta$ value are the best-fit
values for the PSPC image, and the larger $\beta$ is the maximum
boundary value for the GIS fit, respectively. We took A1060 for this
study because it has the lowest intensity.  The variations of the
temperature for the 2 models are $0.05$ and $0.01$ keV for $r< 30'$
and $r> 30'$, respectively, on the average. The largest difference in
a single region ($5'\times 5'$) was $0.13$ keV\@.  Therefore, we may
say that the error in the surface brightness model causes a fairly
small systematic error.

\subsection{Point sources}

Possible influence from hard point sources to the temperature map was
investigated. In figure~4c for the Centaurus cluster, a small hot region
is seen at $35'$ south of the cluster center. The extent before taking
the running mean is $\sim 2 \times 2$ cells ($5'\times 5'$), which is
comparable to the point source image with the XRT response.  This peak
is barely seen in the 0.7 -- 7 keV band, however, the hard band image
above 2 keV shows a clear peak. The position is identified as CCC 032
(Jerjen and Dressler\ 1997); a galaxy of morphological type Im at a
redshift $z=0.00913$.  The temperature derived with an
absorbed MEKAL model for the spectrum within a radius of $5'$ is about 7
keV, consistent with the indication of the temperature map.  The
absorbed X-ray flux in 0.5 -- 10 keV is estimated to be $2.4\times
10^{-12}\ {\rm erg\ cm^{-2}\ s^{-1}}$.  However, about 70\% of the flux is
thought to be the cluster component according to the excess in the
surface brightness.  Therefore, the flux of about $7\times 10^{-13}$ erg
cm$^{-2}$ s$^{-1}$ may be originated from the point source.  

On the other hand, a compact group of galaxies HCG~48 at $29'$ off from
the A1060 center shows no significant structure in the temperature
map, figure~4a. In fact, a spectral fit gives a temperature of 3.3 keV,
consistent with the A1060 ICM\@. We note that the X-ray emission peak
does not correspond to the central elliptical galaxy in HCG~48, but to
a Sa galaxy, ESO~501--G~059 at $2.6'$ south.

These results show that we can distinguish contaminations of point
sources or galaxy groups with angular sizes $\sim 5'$ in terms of
either distinct features in the temperature map or excess emission in
the X-ray image.  Since the extent of the hot region in Centaurus is
considerably larger than the distinguishable size, it is unlikely that
the temperature structures of the ICM are strongly affected by source
contaminations.

\section{Discussion} 

The present {\rm ASCA} mapping observations have revealed significantly
different temperature features in the nearby low-temperature ($ kT
\ltsim 4$ keV) clusters which show fairly symmetric ICM distribution.
The difference is probably resulted from the past history of mergers.
An interesting point is that there appear both hot and cool regions, as
clearly recognized in the Centaurus cluster. Therefore, merging
processes not only cause heating but probably bring in cool gas
originally belonging to infalling subclusters to the ICM\@.

We have confirmed that the ICM of A1060 is almost isothermal with
$kT\sim3.3$ keV over the whole cluster region ($r \sim 1^\circ$).  This
suggests that A1060 has not experienced mergers during the recent a few
Gyr. The metal abundance distribution is also uniform with a rather low
value of $0.3$ solar compared with the other 2 clusters.  The constancy
of metal abundance, as well as the constant temperature, suggests that
the ICM in A1060 had been well mixed after an early starburst phase of
$z\sim 1 - 2$ (Madau\ 1997) where metals have been injected into the ICM\@.

The cD galaxy of A1060, NGC~3311, is smaller than those in typical
nearby bright clusters, and accordingly the central peak of the surface
brightness accompanying the cool component is the weakest among the 3
clusters.  The time scale of radiative cooling at the cluster center is
calculated from the central density of $3\times10^{-3}\ {\rm cm}^{-3}$
and the cooling function (Sutherland and Dopita\ 1993) to be about 4
Gyr, and the central cool gas would take $\sim 10$ Gyr or more to build
up substantially.  These features suggest that the formation process of
A1060 has been rather slow.  One reason could be that the number density
of galaxies in the surrounding region is low. There is an empty space of
about 60 -- 70 Mpc extent in front and behind A1060 along the line of
sight (Richter et al.\ 1982). The low matter density and the early
mixing of the gas may have jointly acted to hold the growth speed of
A1060.

As for AWM~7, the second system studied here, the central $r<20'$ ($\sim
600$ kpc) region shows an isothermal ICM with $kT\sim 3.9$ keV\@.  There
are small temperature deviations near the north and south edges of the
cluster with a low significance of about $1.8\sigma$ (see
table~3) between the hot and cool regions.  The north region (\#1) shows
a cooler temperature than the central region, and vice versa for the
south region. Except for these north and south structures, the constant
temperature suggests that mergers have not occurred for a few Gyr.  In
contrast to A1060, a large-scale metallicity gradient has been found in
AWM~7 (Ezawa et al.\ 1997). A possible scenario based on these results
is that the isothermality was already reached before the early starburst
phase.  If a gas mixing occurred recently, the metal distribution would
become uniform and no more traces the galaxy distribution (see Metzler
and Evrard 1994, Ezawa et al.\ 1997). This excludes the recent gas
mixing in AWM~7, and the isothermality of the ICM also suggests no
recent mergers for a few Gyr.  The calculated cooling time from the
central density of $0.01\ {\rm cm}^{-3}$, $\sim 1$ Gyr, is reasonably
short and can explain that a significant amount of the cool gas has been
accumulated during the prolonged non-merger period.

Another distinctive feature of AWM~7 is the elongated X-ray image in
the east-west direction. This elongation is nearly parallel to the
large-scale filament of the Pisces-Perseus supercluster. There is,
however, no significant temperature structure along this direction as
seen in figure~4b.  Tidal deformation and directional mass accretion
have been considered as possible origins of the elongated morphology
(Neumann and B\"ohringer 1995). The tidal force from the Perseus
cluster ($2\times 10^{15} M_\odot$) was shown to deform the gas by
only 2\%, which is a factor of 5--10 too weak to cause the observed
ellipticity.  On the other hand, if the directional mass accretion
takes the form of subcluster mergers, we would expect a clear
temperature structure along the east-west direction. Since the tidal
deformation itself would not cause large temperature variation, the
constant temperature in AWM~7 suggests that the underlying dark matter
may have an extended structure along the filament of the
Pisces-Perseus supercluster.

The most remarkable result in the present observations is the
temperature distribution of the Centaurus cluster. We have shown that
the hot region reported by Churazov et al.\ (1999) is more extended
and also covers the cluster center, which is seen after subtracting
the cool component (figure 4d). Also, there are cool regions in the
northeast and northwest near the cluster edge.  The spectral fits show
that the temperatures in the hot and cool regions differ by a factor
of about 2, and the deviations from the surrounding values are $\pm
20-30\%$. 

Since both the central and the southeast hot regions in the Centaurus
show similar temperatures, $kT \sim 4.3$ keV, they may have been
heated in a single major merger. The radiative cooling time at the
cluster center with an electron density of $0.02\ {\rm cm}^{-3}$ is
about 0.5 Gyr, much less than the time for the temperature variation
to be smoothed out. Therefore, the central cool component may have
been accumulated recently, after the merger which have probably
occurred a few Gyr ago.  However, there is strong metal concentration
in both the hot and the cool components (Fukazawa et al.\ 1994). The
recent accumulation after the gas mixing due to the merger cannot
explain this metallicity gradient, and the cool gas is more likely to
have been built up at least before the last merger which has caused
the temperature variation.

The stellar mass within the central 50 kpc of the Centaurus cluster is
estimated to be $2 \times 10^{12} M_\odot$ (Ikebe et al.\ 1999).  Such
a mass concentration, i.e.\ the formation of the cD galaxy, must have
started in a very early time. This suggests that the strong cool
component has been established all the way through the early starburst
phase until the beginning of the major merger. In this case, the
central cool component was probably massive enough to survive the
merger which have caused a shock heating and raised temperature by
nearly 1 keV\@.  Recently, G\'omez et al.\ (2000) performed numerical
simulations of head-on mergers in cooling-flow clusters. In cases
where the initial cooling times are very small, even though the flows
are once disrupted, the central cooling time remains less than the
Hubble time and the flows are re-established in a few Gyr. These
clusters would naturally be classified as cooling-flow clusters even
though they have experienced significant mergers. Regardless of the
origin of the cool component, the Centaurus cluster may correspond to
these cases of survived cool components.

\section{Conclusion}

Both A1060 and AWM~7 show isothermal ICM for the most part of the
cluster region, and suggest that they have not experienced merging
processes for recent a few Gyr. The difference in the metal abundance
distribution imply that the gas mixing due to a merger occurred before
and after the metal enrichment of the ICM in AWM~7 and A1060,
respectively.  Absence of a significant temperature structure along
the elongation of the X-ray morphology of AWM~7 suggests a possibility
that a tidal deformation by an unknown large gravitational potential
may be working.  The Centaurus cluster exhibits remarkable temperature
structures characterized by the extended hot and outer cool regions.
A shock heating due to a subcluster merger into the main gravitational
potential can consistently explain the observed temperature increase
by about 20--30\%.  It is likely that the central cool component was
already established at the time of the merger, because the cool
component shows strong metal concentration around the cD galaxy.
Therefore, the central cool component in Centaurus was not disrupted
by the major merger which has created the complicated temperature
structure.

\bigskip We appreciate the ASCA\_ANL, SimASCA, and SimARF teams for
the support of data analysis.  We also thank Drs.\ S. Sasaki,
M. Watanabe, and M.  Takizawa for stimulating discussion and useful
comments.  This work was partly supported by the Grants-in Aid for
Scientific Research No.\ 08404010 and No.\ 12304009 from the Japan
Society for the Promotion of Science.

\section*{References} 

\re Anders E., Grevesse N. 1989, Geochim. Cosmochim. Acta 53, 197
\re Allen S.W., Fabian A.C.\ 1994, MNRAS, 269, 409 
\re B\"{o}hringer H.\ 1994, in Cosmological Aspects of X-Ray Clusters of Galaxies, ed W.C.\ Seitter (Kluwer Academic Publishers, Dordrecht, Boston, and London) v441, p123
\re Churazov E., Gilfanov M., Forman W., Jones C.\ 1999, ApJ, 520, 105 
\re Donnelly R.H., Markevitch M., Forman W., Jones C., Churazov E., Gilfanov M.\ 1999, 513, 690 
\re Ezawa H., Fukazawa, Y., Makishima, K., Ohashi, T., Takahara, F., Xu, H., Yamasaki, N.Y.\ 1997, ApJ, 490, L33 
\re Fabian A.C. 1994, ARA\&A, 32, 227 
\re Fukazawa, Y., Ohashi, T., Fabian, A. C., Canizares, C. R., Ikebe, Y., Makishima, K., Mushotzky, R. F., Yamashita, K. 1994, PASJ, 46, L55
\re G\'omez P. L., Loken C., Roettiger K., Burns J. O. 2000 astro-ph/0009465 
\re Honda H., Hirayama M., Watanabe M., Kunieda H., Tawara Y., Yamashita K., Ohashi T., Hughes J.P., Henry J.P. 1996, ApJ 473, L71 
\re Ikebe Y. 1995, PhD Thesis, The University of Tokyo
\re Ikebe Y., Makishima K., Fukazawa Y., Tamura T., Xu H., Ohashi T., Matsushita K.\ 1999, ApJ, 525, 58 
\re Ishisaki, Y.\ 1996, PhD Thesis, The University of Tokyo 
\re Jerjen H., Dressler A.\ 1997, A\&AS, 124, 1 
\re Kikuchi K., Itoh C., Kushino A., Furusho T., Matsushita K., Yamasaki N.Y., Ohashi T., Fukazawa, Y.\ et al.\ 2000, ApJL, 531, 95 
\re Madau P.\ 1997, in Star Formation Near and Far, ed S.S.\ Holt, L.G.\ Mundy (Woodbury, New York) 393, p481  
\re Makishima K., Tashiro M., Ebisawa K., Ezawa H., Fukazawa Y., Gunji S., Hirayama M., Idesawa E.\ et al.\ 1996, PASJ 48, 171 
\re Markevitch M., Vikhlinin A.\ 1997, ApJ, 474, 84 
\re Markevitch M., Forman W.R.,  Sarazin C.L., Vikhlinin A.\ 1998, ApJ, 503, 77 
\re Metzler C.A., Evrard A.E.\ 1994, ApJ, 437, 564 
\re Neumann D.M., B\"{o}hringer H.\ 1995, A\&A, 301, 865 
\re Ohashi T., Ebisawa K. Fukazawa Y., Hiyoshi K., Horii M., Ikebe Y.,  Ikeda H., Inoue H.\ et al.\ 1996, PASJ 48, 157 
\re Richter O.-G., Materne J., Huchtmeier W.K.\ 1982, A\&A, 111, 193 
\re Schindler S., M\"{u}ller E.\ 1993, A\&A, 272, 137
\re Serlemitsos P.J., Jalota L., Soong Y., Kunieda H., Tawara Y., Tsusaka Y., Suzuki H., Sakima Y.\ et al.\ 1995, PASJ 47, 105
\re Shibata R., Matsushita K., Yamasaki N.Y., Ohashi T., Ishida M., Kikuchi K., B\"ohringer H., Matsumoto H.\ 2001, ApJ, 549, in press (astro-ph/0010380) 
\re Snowden S.L., McCammon D., Burrows D.N., Mendenhall J.A.\ 1994, ApJ, 424, 714 
\re Stark A.A., Gammie C.F., Wilson R.W., Bally J., Linke R.\ 1992, ApJS, 79, 77 
\re Sutherland R.S., Dopita M.A.\ 1993, ApJS, 88, 253 
\re Tamura T., Day C.S., Fukazawa Y., Hatsukade I., Ikebe Y., Makishima K., Mushotzky R.F., Ohashi T.\ et al.\ 1996, PASJ, 48, 671 
\re Tamura T., Ikebe Y., Fukazawa Y., Makishima K., Ohashi T.\ 1997, in X-ray Imaging and Spectroscopy of Cosmic Hot Plasmas, ed F.\ Makino, K.\ Mitsuda (Universal Academy Press, Tokyo) p127
\re Tamura T., Makishima K., Fukazawa Y., Ikebe Y., Xu H.\ 2000, ApJ, 535, 602 
\re Tanaka Y., Inoue H., Holt S.S.\ 1994, PASJ 46, L37 
\re Watanabe M., Yamashita K., Furuzawa A., Kunieda H., Tawara Y. 1999, ApJ, 527, 80 
\re Xu, H. et al.\ 1997, PASJ, 49, 9 


\begin{table*}[htb]
\begin{center}
  Table~1.\hspace{4pt} Log of ASCA observations
\end{center}
{\small
\begin{tabular*}{\textwidth}{@{\hspace{\tabcolsep}
    \extracolsep{\fill}}lclrrc}\hline\hline\\[-6pt]
Cluster  & Sequence No. &\multicolumn{1}{c}{Date} & 
\multicolumn{2}{c}{Coordinates (J2000)} & Exposure $^{\dagger}$\\
         &          &           & \multicolumn{1}{c}{R.A.}&
\multicolumn{1}{c}{Decl.}&\multicolumn{1}{c}{(s)}\\\hline
A1060   
& 80004000 & 93 Jun 28 & 159.42 & $-27.54$\hspace*{5mm} & 21940\hspace*{5mm}\\
& 83024000 & 94 Dec 12 & 159.17 & $-27.08$\hspace*{5mm} & ~6582\hspace*{5mm}\\
& 83024010 & 94 Dec 19 & 159.17 & $-27.05$\hspace*{5mm} & 10451\hspace*{5mm}\\
& 87009000 & 99 Dec 19 & 158.53 & $-27.23$\hspace*{5mm} & 14129\hspace*{5mm}\\
& 87010000 & 99 Dec 20 & 159.27 & $-27.64$\hspace*{5mm} & 16385\hspace*{5mm}\\
& 87011000 & 99 Dec 21 & 158.85 & $-27.96$\hspace*{5mm} & 10448\hspace*{5mm}\\
& 87012000 & 99 Dec 21 & 158.39 & $-27.69$\hspace*{5mm} & 15502\hspace*{5mm}\\
\hline AWM~7    
& 80036000 & 93 Aug ~7  & 43.46  & $41.42$\hspace*{5mm}  & ~9392\hspace*{5mm}\\
& 80037000 & 93 Aug ~7  & 43.06  & $41.40$\hspace*{5mm}  & 14596\hspace*{5mm}\\
& 81011000 & 94 Feb 10 & 44.27  & $41.78$\hspace*{5mm}  &  ~7525\hspace*{5mm}\\
& 81011010 & 94 Feb 10 & 43.32  & $41.78$\hspace*{5mm}  & 13632\hspace*{5mm}\\
& 81011020 & 94 Feb 11 & 43.88  & $42.20$\hspace*{5mm}  & 12136\hspace*{5mm}\\
& 81011030 & 94 Feb 11 & 43.88  & $41.62$\hspace*{5mm}  & 10189\hspace*{5mm}\\
\hline Centaurus
& 80032000 & 93 Jun 30 & 192.47 & $-41.23$\hspace*{5mm} & 11477\hspace*{5mm}\\
& 80033000 & 93 Jul ~~5 & 192.55 & $-40.96$\hspace*{5mm} & 11326\hspace*{5mm}\\
& 80034000 & 93 Jul ~~5 & 192.79 & $-41.32$\hspace*{5mm} &  ~8999\hspace*{5mm}\\
& 83026000 & 95 Jul \,19 & 192.46 & $-41.37$\hspace*{5mm} & 45899\hspace*{5mm}\\
& 86000000 & 98 Feb ~3  & 191.98 & $-40.58$\hspace*{5mm} &  ~6777\hspace*{5mm}\\
& 86001000 & 98 Feb ~3  & 191.28 & $-40.78$\hspace*{5mm} & 10580\hspace*{5mm}\\
& 86002000 & 98 Feb ~4  & 191.03 & $-41.30$\hspace*{5mm} &  ~7656\hspace*{5mm}\\
& 86003000 & 98 Feb ~4  & 191.31 & $-41.77$\hspace*{5mm} &  ~9634\hspace*{5mm}\\
& 86004000 & 98 Feb ~4  & 191.97 & $-41.87$\hspace*{5mm} &  ~8206\hspace*{5mm}\\
& 86005000 & 98 Feb ~5  & 192.50 & $-41.56$\hspace*{5mm} &  ~7706\hspace*{5mm}\\
& 86006000 & 98 Feb ~5  & 192.50 & $-41.01$\hspace*{5mm} & 11541\hspace*{5mm}\\
         \hline
       \end{tabular*}
\vspace{6pt}\par\noindent
$\dagger$
The net exposure time after all the screening and cleaning procedures.}
\end{table*}

\clearpage

\begin{table*}[htb]
\begin{center}
  Table~2.\hspace{4pt} Spectral model parameters and sequence numbers
  of PSPC observations of each cluster for the ray-tracing
  simulations.
\end{center}
\begin{tabular*}{\textwidth}{@{\hspace{\tabcolsep}
    \extracolsep{\fill}}lcccc}\hline\hline\\[-6pt]
Cluster  & $N_{\rm H}$ & $kT$ & $Z$ & PSPC observations\\
         & ($10^{20}\ {\rm cm}^{-2}$) & (keV) & (solar)& Sequence no.\\\hline
A1060   & 3.5 & 3.3 & 0.31 & 800200\\
AWM~7    & 8.7 & 3.6 & 0.51--0.16 & 800168\\
Centaurus& 8.8 & 4.0 & 1.5--0.2  & 800192, 800321, 800322, 800323, 800324\\
         \hline
       \end{tabular*}
\vspace{6pt}\par\noindent
\end{table*}


\begin{table*}[htb]
\begin{center}
  Table~3.\hspace{4pt} Results of the spectral fits for selected
  regions. The errors are statistical and 90\%-confidence levels for
  single parameter of interest.
\end{center}
\begin{tabular*}{\textwidth}{@{\hspace{\tabcolsep}
    \extracolsep{\fill}}llcccc}\hline\hline\\[-6pt]
Cluster   & region no.$^\dagger$ & $N_{\rm H}$ &  $kT$             & $Z$     & $\chi^2/d.o.f$ \\
          & & ($10^{20}\ {\rm cm}^{-2}$) &  (keV)            & (solar) &                \\\hline 
A1060    & \#1 (med)  & 3.5(fix)             & 3.00$^{+0.14}_{-0.13}$ & 0.44$^{+0.14}_{-0.12}$ & 67.0/55\\ 
          & \#2 (hot)  & 3.5(fix)             & 3.19$^{+0.19}_{-0.18}$ & 0.30$^{+0.15}_{-0.13}$ & 60.1/55\\
AWM~7     & \#1 (cool) & 8.7(fix)             & 3.60$^{+0.60}_{-0.44}$ & 0.19$^{+0.38}_{-0.19}$ & 22.5/18\\
          & \#2 (hot)  & 8.7(fix)             & 4.61$^{+1.16}_{-0.70}$ & 0.35$^{+0.58}_{-0.23}$ & 24.2/18\\
Centaurus & \#1 (med)  & 8.8(fix)             & 3.54$^{+0.09}_{-0.09}$ & 0.41$^{+0.07}_{-0.07}$ & 47.8/45\\
          &  \#2 (hot) & 8.8(fix)             & 4.31$^{+0.17}_{-0.17}$ & 0.32$^{+0.09}_{-0.08}$ & 60.5/45 \\
          & \#3 (cool) & 8.8(fix)             & 2.24$^{+0.35}_{-0.30}$ & 0.08$^{+0.41}_{-0.08}$ & 11.1/14\\ 
         \hline
       \end{tabular*}
\vspace{6pt}\par\noindent
$\dagger$ Corresponding regions are shown in figure 6--8.

\end{table*}

\clearpage

\begin{fv}{1}{0.1cm}{The GIS images of (a) A1060, (b) AWM~7, and 
(c) the Centaurus cluster in the energy range 0.7 -- 7 keV\@. Circles
with a radius of $22'$ show the GIS field of view. The images are
smoothed by a Gaussian function with $\sigma = 1'$. The background is
subtracted, and corrections for exposure time are performed.}\\
\end{fv} 

\begin{fv}{2}{0.1cm}
{Fractions of the photon origin estimated from the ray-tracing
simulation of the multi-pointing observations of the 3 clusters. The
top curves indicate the flux coming from the observed region, and the
curves in the middle and bottom show the contaminating fraction from
inner and outer regions, respectively. The integrated radius of each
annulus is shown as the thick bars only for A1060.  }\\
\centerline{\psfig{figure=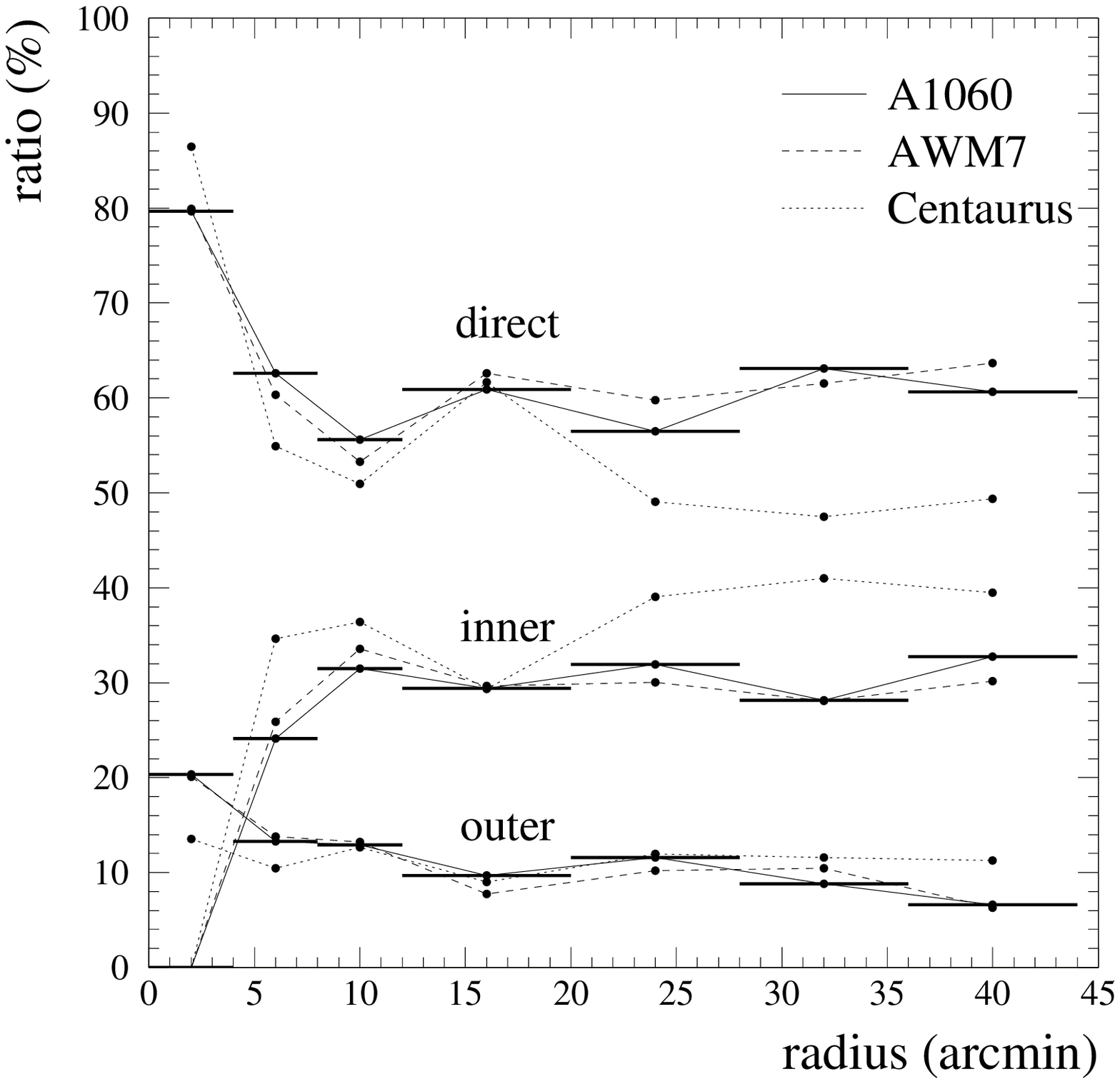,width=8cm}}
\end{fv} 

\begin{fv}{3}{0.1cm} {Variation of the {\it HR} in the azimuthal direction
    with radius $10'-20'$ for the three clusters. The filled circles
    show the observed values, and the solid lines show the simulated
    values for isothermal clusters.  The error bars show 90\%
    confidence levels.}\\
    \centerline{\psfig{figure=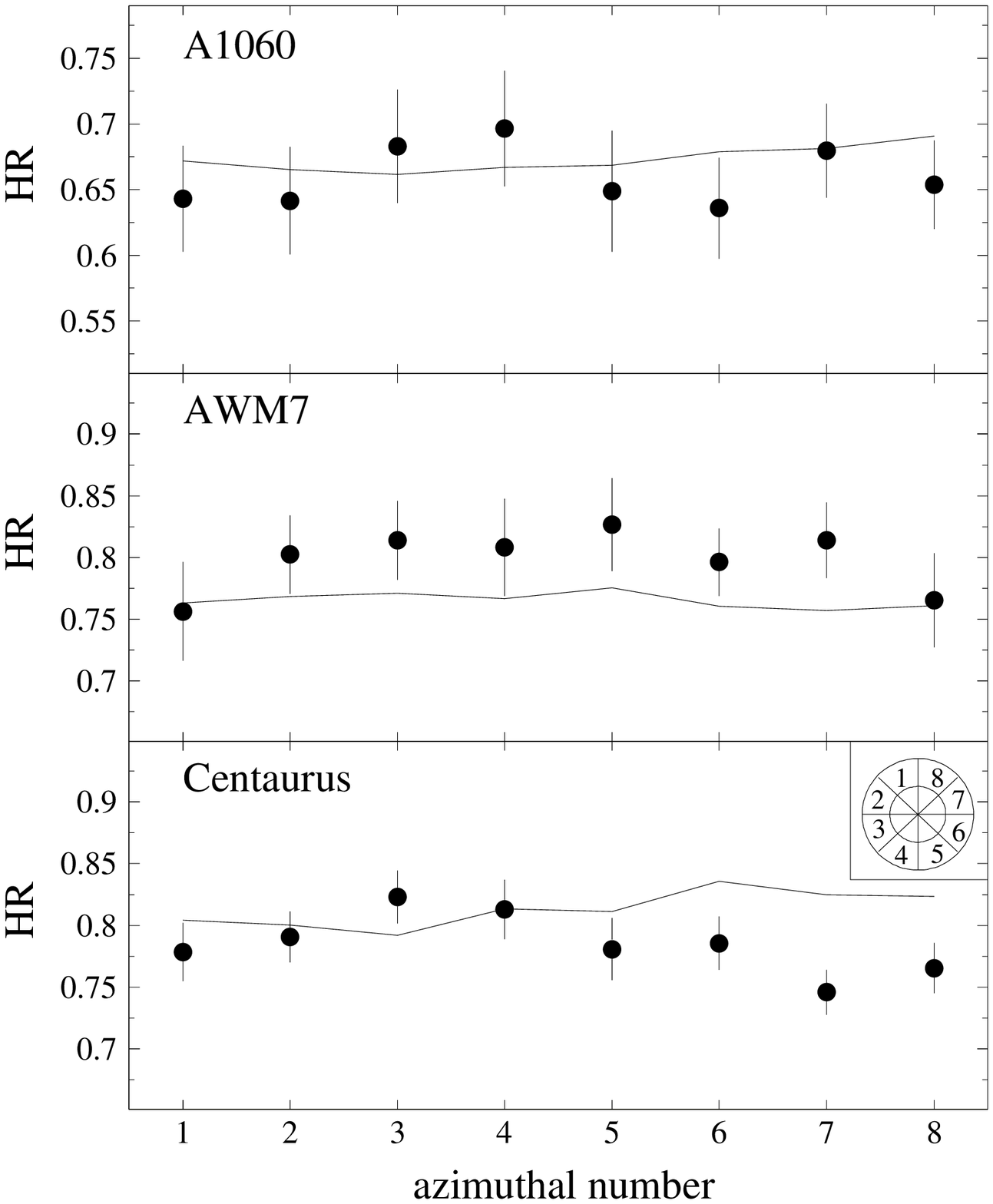,width=8cm}}
\end{fv}

\begin{fv}{4-a,b}{0.1cm}
{Color-coded temperature maps for (a) A1060 and (b) AWM~7. The
contours indicate X-ray intensity measured with the GIS\@. }
\end{fv}
\begin{fv}{4-c,d}{0.1cm}
{Color-coded temperature maps for the Centaurus cluster. The panel (c)
shows results without correction for the central cool component. In
the panel (d), the cool component was subtracted from the data based
on the radial profile given by Ikebe et al.\ (1999). }
\end{fv}

\begin{fv}{5-a,b}{0.1cm}
  {Azimuthal profiles of temperature in 7 annular regions for (a)
    A1060 and (b) AWM~7. The ring radii are indicated in the abscissa
    and each ring is divided to 8 azimuthal sections with an opening
    angle of $45^\circ$ (4 sections with $90^\circ$ for the innermost
    ring of $2'-4'$) in a counterclockwise direction from the north.
    The error bars show 90\% confidence levels. The solid lines show
    the mean temperatures by constant-temperature fits.}
    \centerline{\psfig{figure=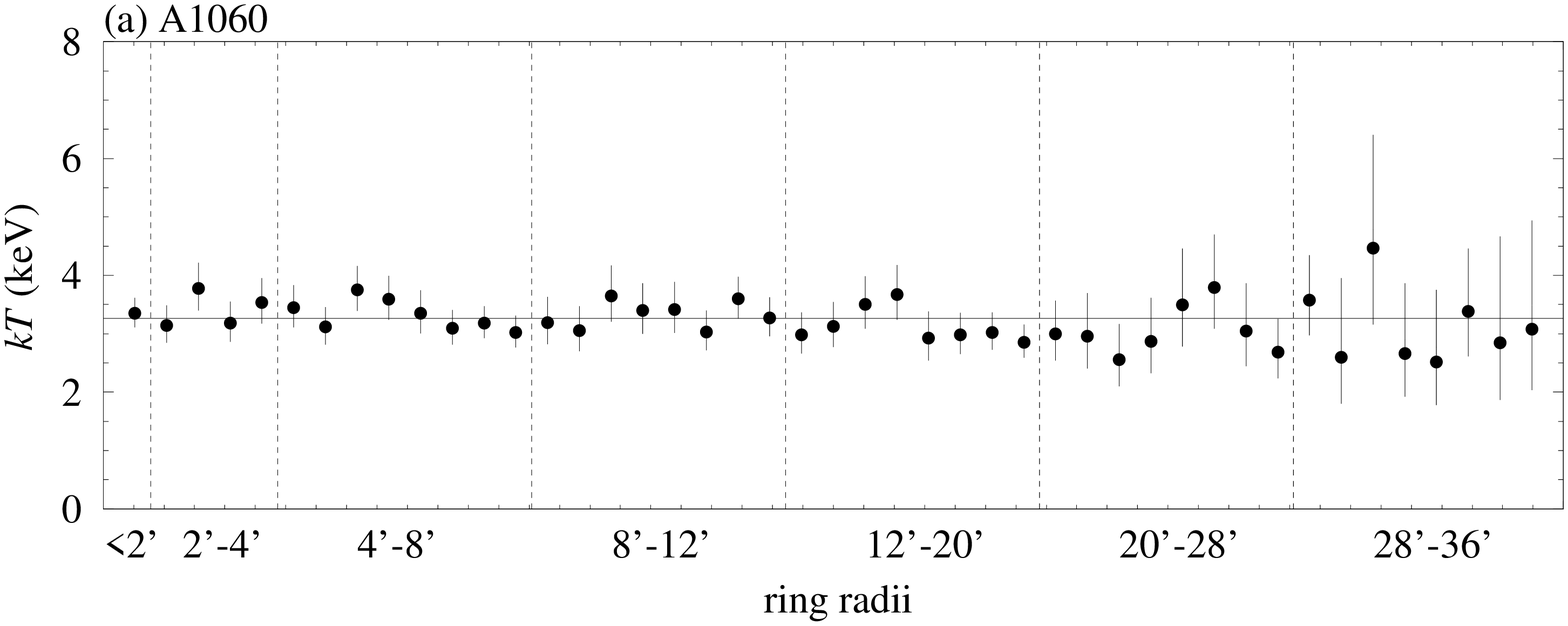,width=11cm}}
    \centerline{\psfig{figure=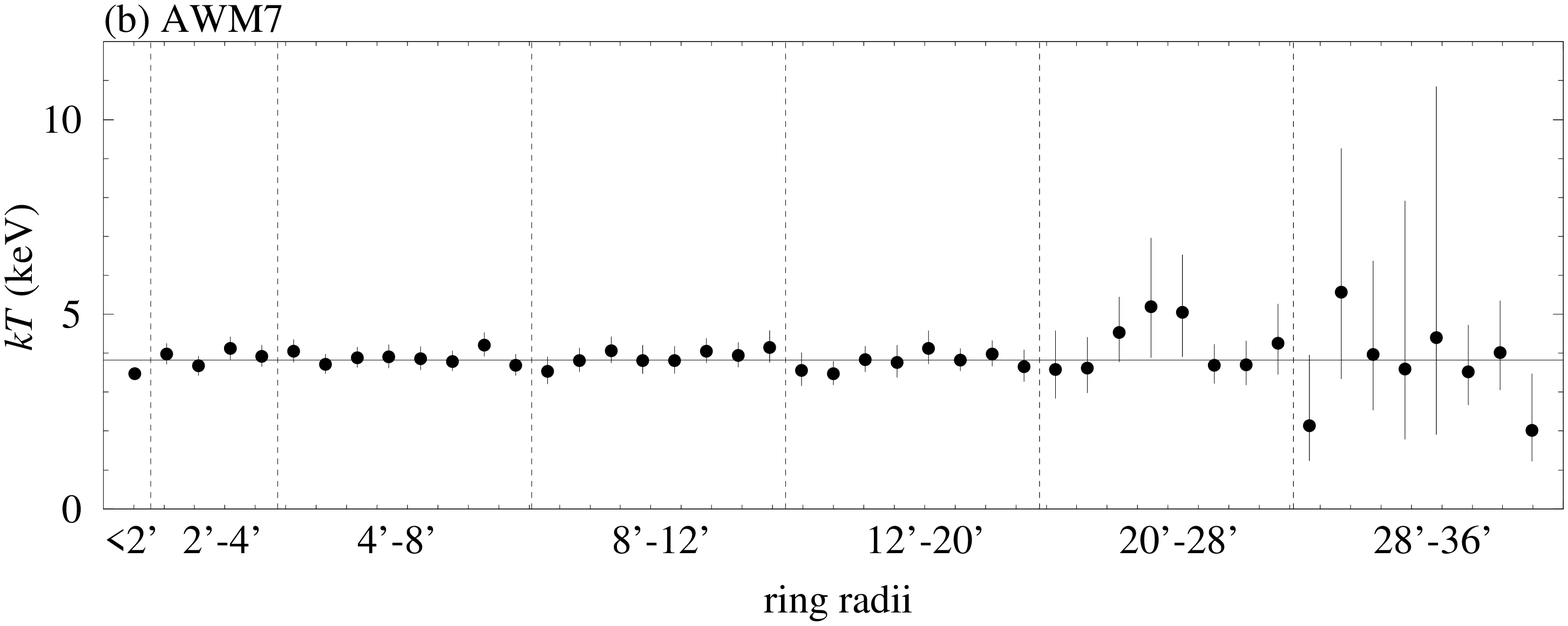,width=11cm}}
\end{fv}
\begin{fv}{5-c,d}{0.1cm}
  {Azimuthal temperature profiles for the Centaurus cluster, similar
    to the plots in figures 5(a) and (b). Figure~5 (c) and (d) are
    before and after the correction for the central cool component,
    respectively. The error bars in (d) are increased by including the
    systematic error of the cool-component intensity by 10\%.} 
    \centerline{\psfig{figure=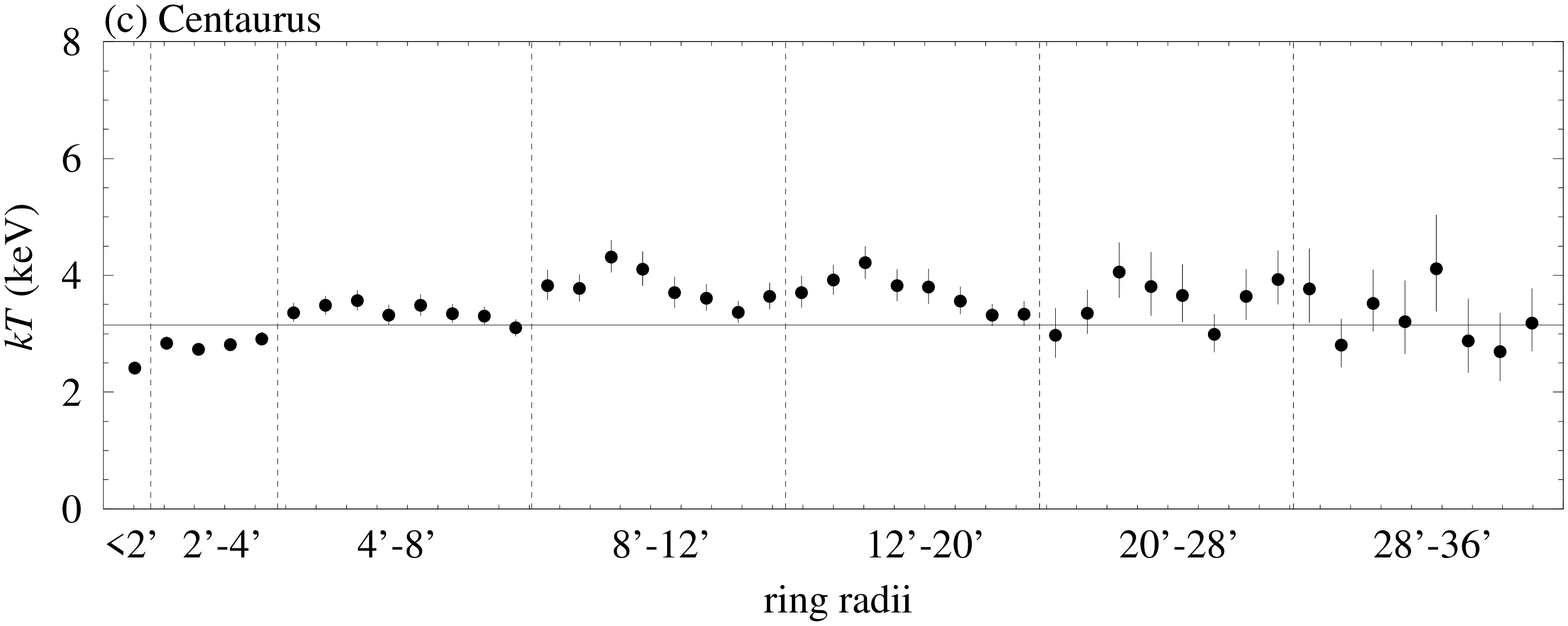,width=11cm}}
    \centerline{\psfig{figure=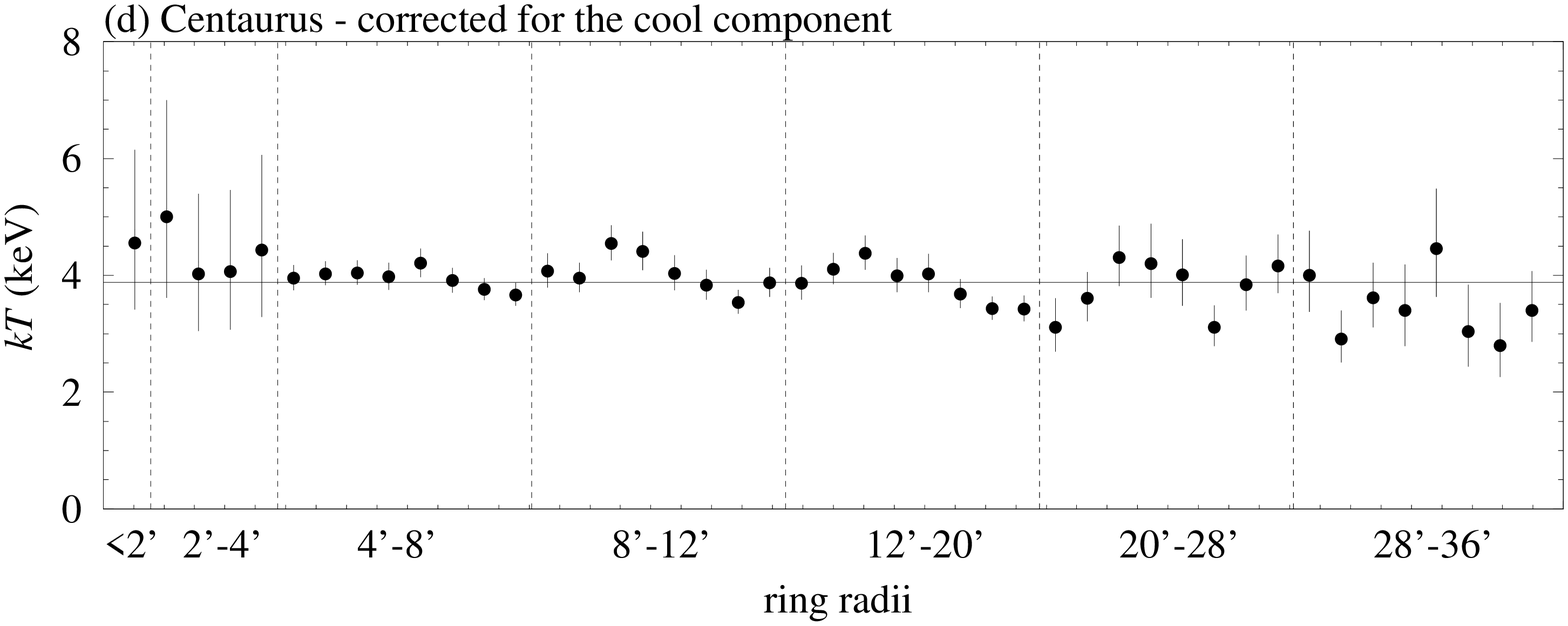,width=11cm}}
\end{fv} 

\clearpage

\begin{fv}{6}{0.1cm}
  {Pulse-height spectra for two selected regions in A1060 fitted with
    thermal models. The bottom panels show data-to-model ratios for a
    common model, which is the best-fit model of \#1 (``med'') shown
    by the thick line in the top panel.  The right panel shows
    corresponding regions, in which the dashed circles show radii of
    $20'$ and $40'$.\\
    \psfig{figure=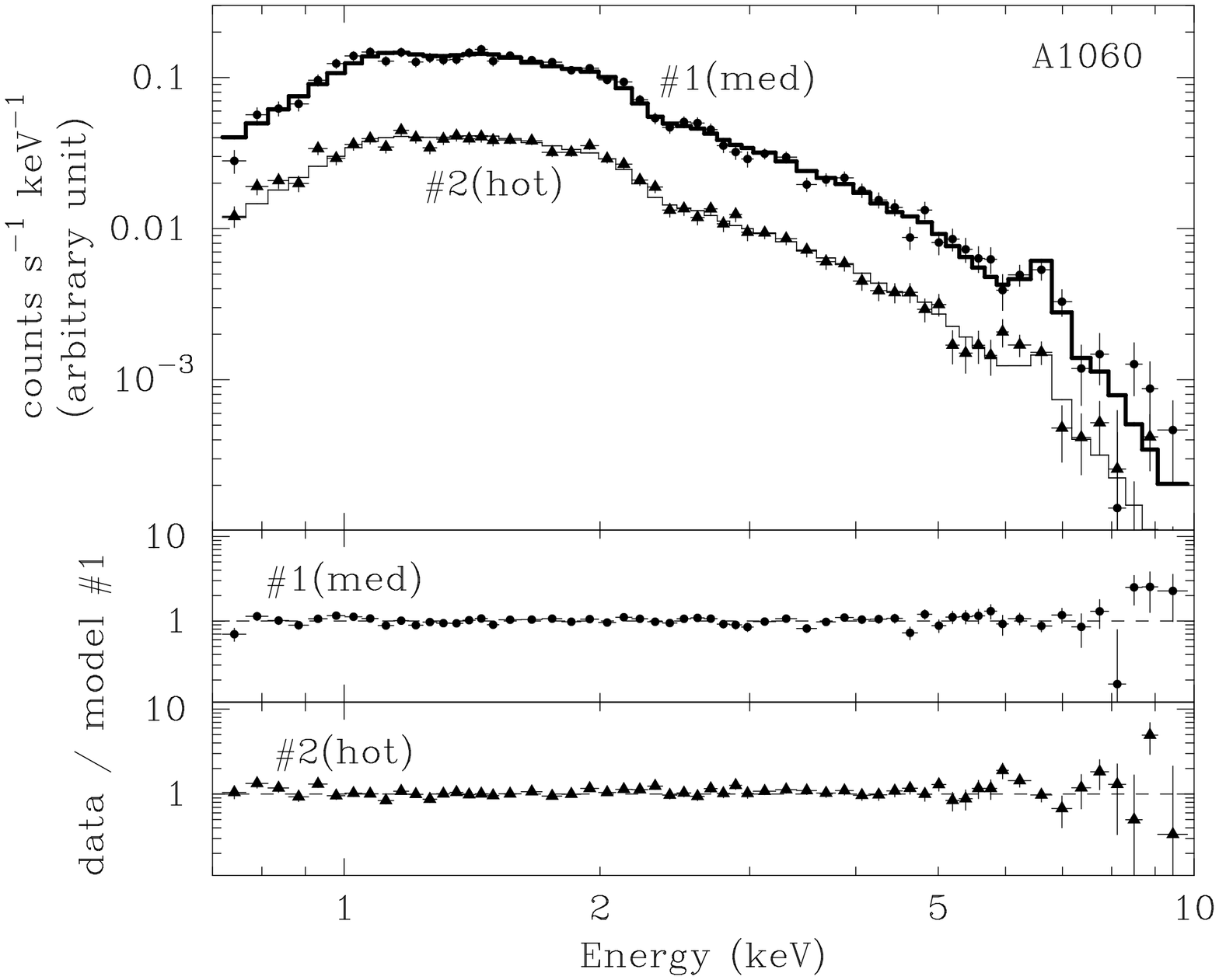,width=6.5cm,angle=-90}
    \psfig{figure=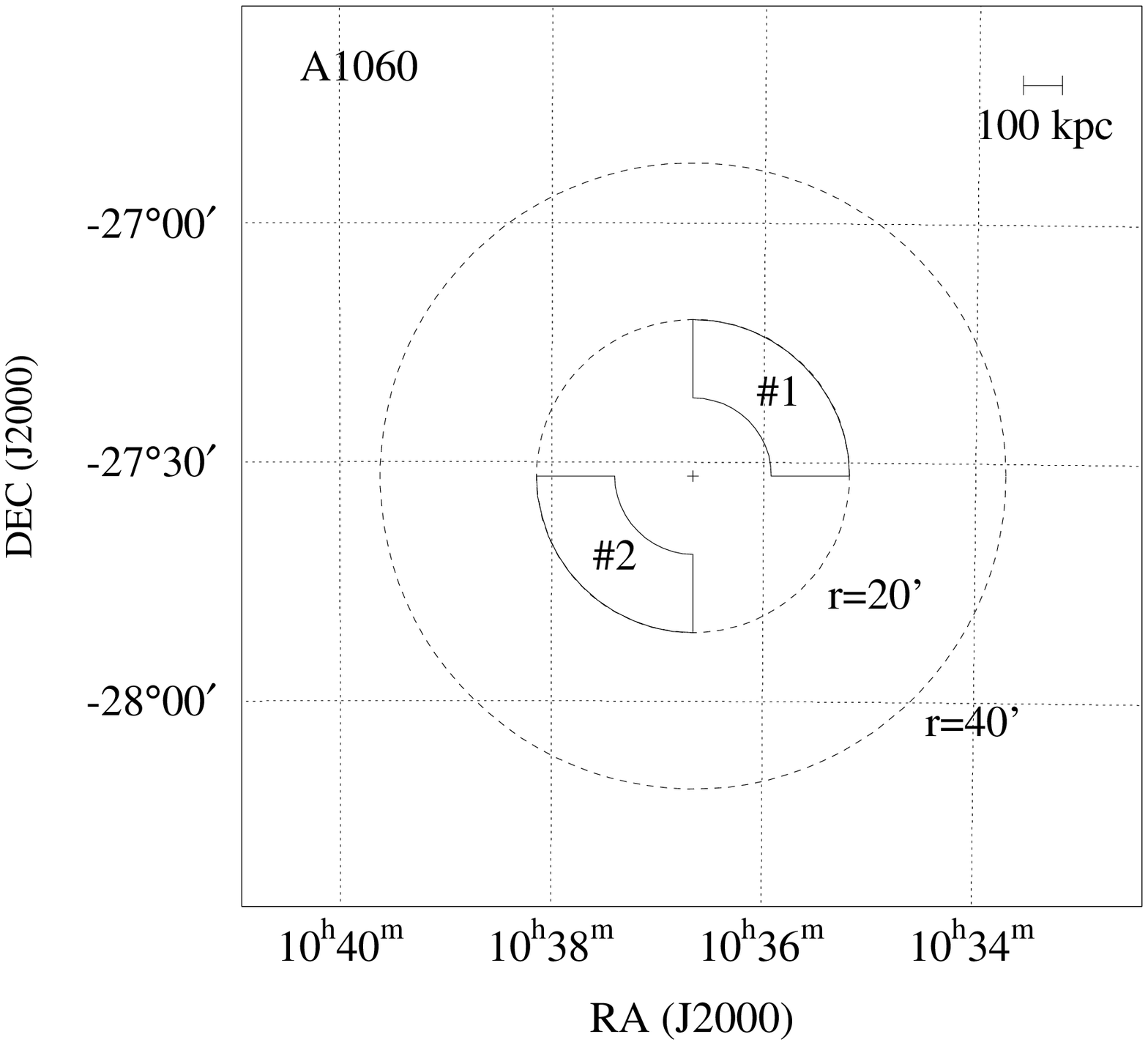,width=6cm}}
\end{fv}

\begin{fv}{7}{0.1cm}
  {Pulse-height spectra for two regions (top panel)in AWM~7, and
    normalized ratios to the best-fit model of \#1 (bottom panel).
    The right panel shows the regions in the same way as in figure~6.\\
    \psfig{figure=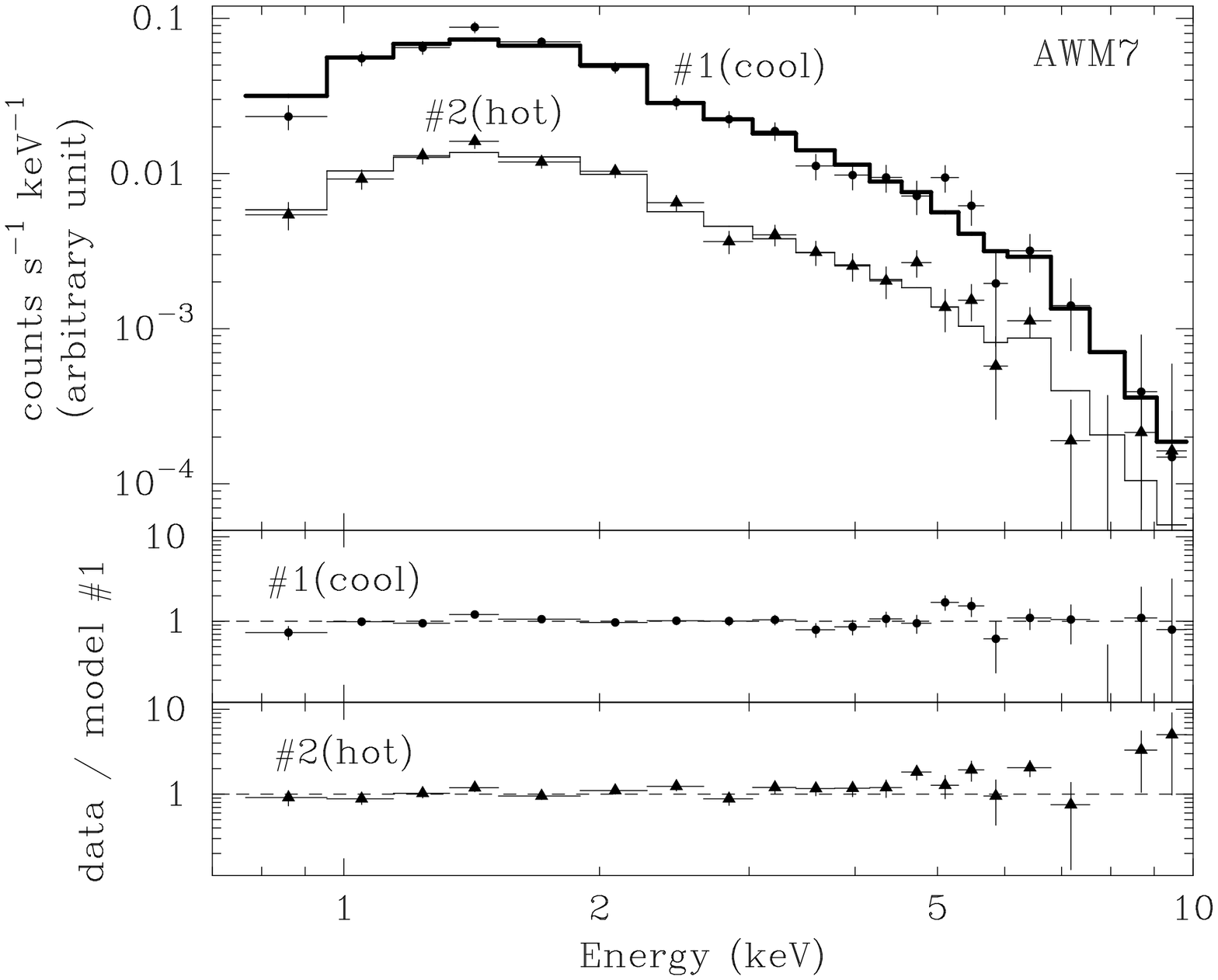,width=6.5cm,angle=-90}
    \psfig{figure=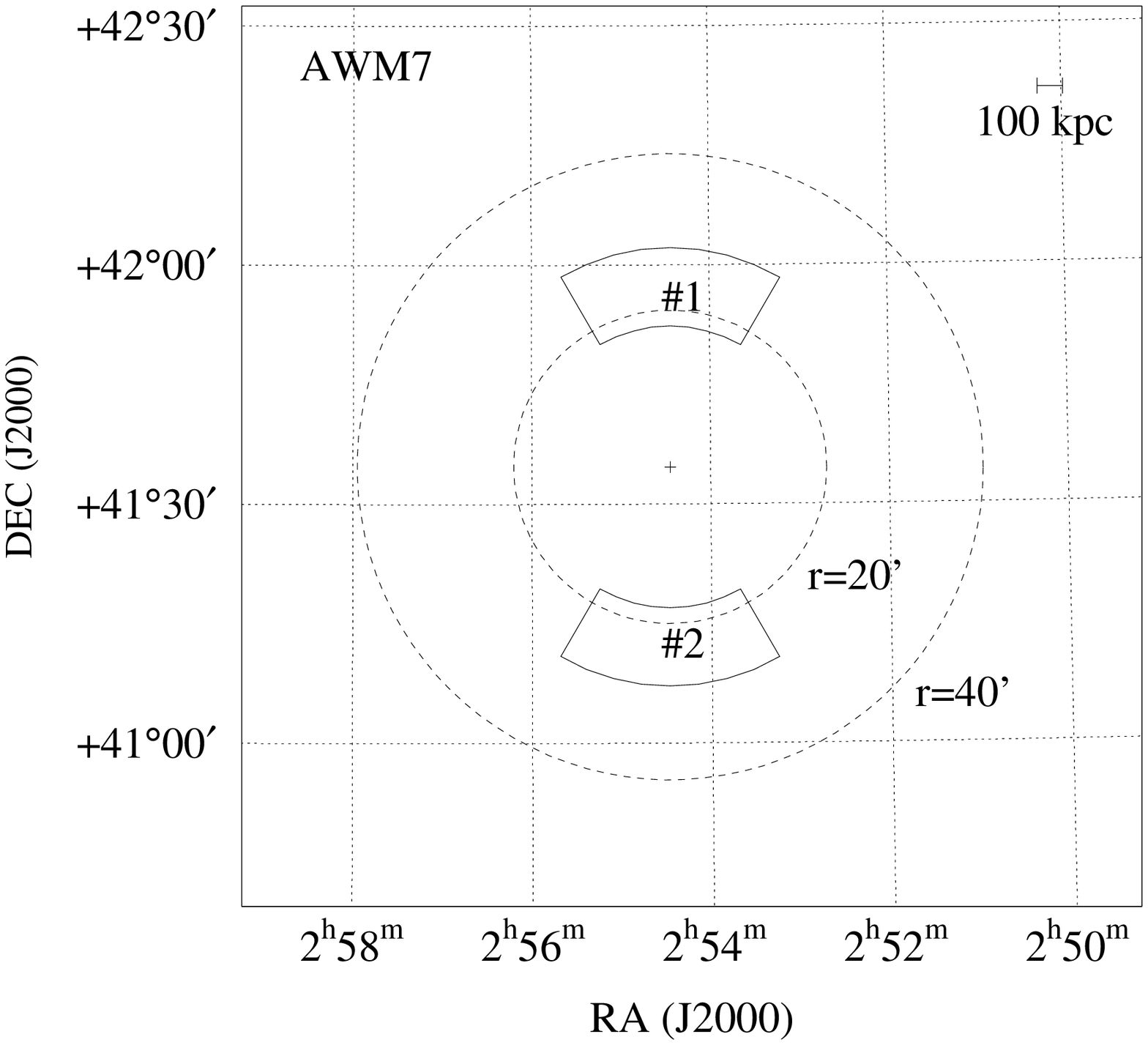,width=6cm}}
\end{fv}

\begin{fv}{8}{0.1cm}
  {Pulse-height spectra (top panel) for the Centaurus cluster,
    normalized ratios to the best-fit model of \#1 (bottom panel), and
    the selected regions (right panel), in the same way as figure~6.\\
    \psfig{figure=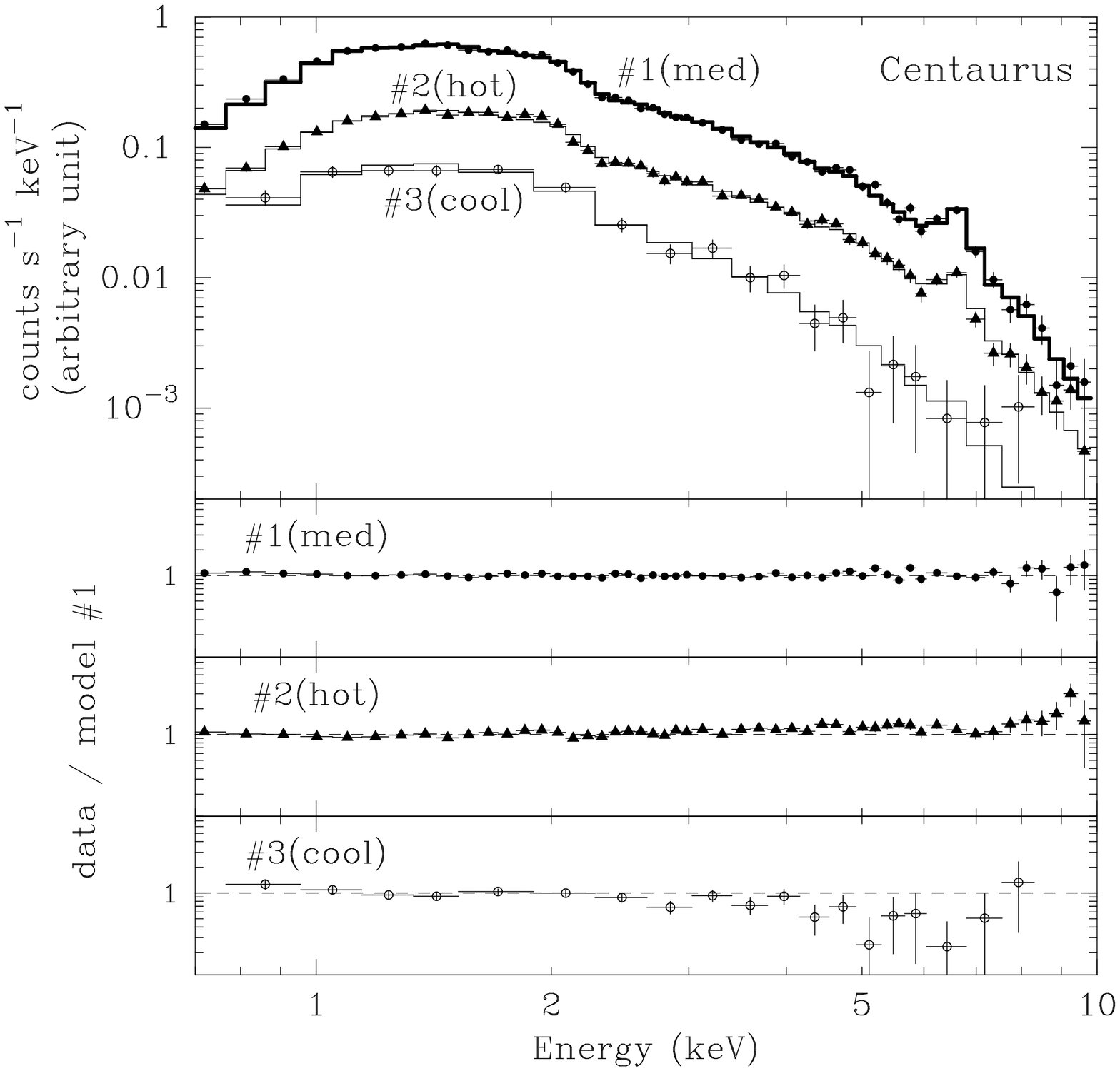,width=6.5cm,angle=-90}
    \psfig{figure=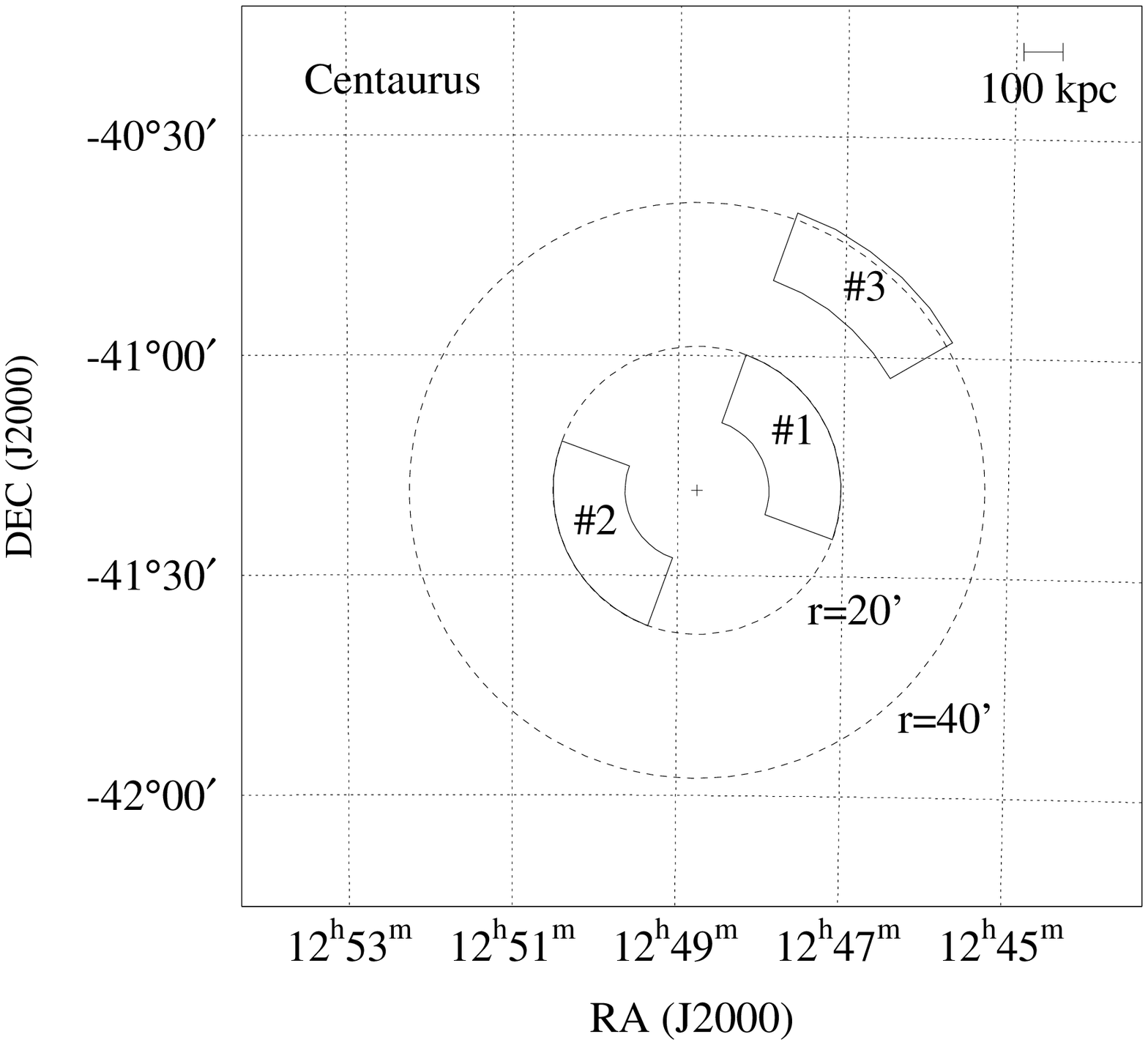,width=6cm}}
\end{fv}

\end{document}